\documentclass[letterpaper,3p,twocolumn,number]{elsarticle}
\usepackage{mathptmx}
\usepackage[T1]{fontenc}
\usepackage[utf8]{inputenc}
\usepackage{listings}
\usepackage{color}
\usepackage{units}
\usepackage{amsmath}
\usepackage{graphicx}
\usepackage[unicode=true, 
 bookmarks=true,bookmarksnumbered=false,bookmarksopen=false,
 breaklinks=false,pdfborder={0 0 0},backref=false,colorlinks=true]
 {hyperref}
\hypersetup{pdftitle={A Probability-Conserving Biasing Technique},
 pdfauthor={Marcus H. Mendenhall and Robert A. Weller },
 pdfsubject={version $Revision: 1.43 $},
 pdfkeywords={Keywords: non-analogous Monte-Carlo, non-Boltzmann tally, variance reduction, Geant4, cross-section biasing, particle splitting, history splitting}}

\makeatletter

\pdfpageheight\paperheight
\pdfpagewidth\paperwidth

\newcommand{\noun}[1]{\textsc{#1}}
\newcommand{\lyxdot}{.}

\def\@journal{NIM A}

\makeatother

\begin{document}

\title{A probability-conserving cross-section biasing mechanism for variance
reduction in Monte Carlo particle transport calculations}

\author{Marcus H. Mendenhall\corref{cor1}}

\ead{marcus.h.mendenhall@vanderbilt.edu}

\author{Robert A. Weller}

\ead{robert.a.weller@vanderbilt.edu}

\cortext[cor1]{ Corresponding Author}

\address{Vanderbilt University Department of Electrical Engineering, P.O.
Box 351824B, Nashville, TN, 37235, USA}
\begin{abstract}
In Monte Carlo particle transport codes, it is often important to
adjust reaction cross sections to reduce the variance of calculations
of relatively rare events, in a technique known as non-analogous Monte
Carlo. We present the theory and sample code for a Geant4 process
which allows the cross section of a G4VDiscreteProcess to be scaled,
while adjusting track weights so as to mitigate the effects of altered
primary beam depletion induced by the cross section change. This makes
it possible to increase the cross section of nuclear reactions by
factors exceeding $10^{4}$ (in appropriate cases), without distorting
the results of energy deposition calculations or coincidence rates.
The procedure is also valid for bias factors less than unity, which
is useful, for example, in problems that involve computation of particle
penetration deep into a target, such as occurs in atmospheric showers
or in shielding.\end{abstract}
\maketitle

\section{Background}

In any Monte-Carlo particle transport calculation, the quality of
the computed result is limited by the statistical uncertainty in the
counting of the random events that are modeled. Often it is interesting
to compute the probability of very rare events, since these may have
particularly large significance in certain applications. A typical
example of relevance to the semiconductor community is single event
effects (SEEs) created by the breakup of a heavy nucleus by an incoming
cosmic ray. Such events may be very rare, but because they can cause
a single incoming particle to create multiple outgoing particles,
often with very high linear energy transfer (LET), they can foil SEE
mitigation techniques which rely on an assumption that only one device
in a region can be upset by an individual event \citep{Weller_bigpaper_2010_5550410,Sheshadri_IRPS_2010_multinode_5488683,Clemens_TNS_2009_fragmentation_3158}.
Rare events can also be generated in a converse situation, in which
the very strong absorption of particles makes it difficult to look
at the types of events that are due to particles that have penetrated
very deeply into the material, as would be the case for events due
to particles that have penetrated shielding. 

To improve the statistical variance in such calculations, it is common
to modify the cross section for potentially interesting processes,
and to appropriately weight the resulting calculation to compensate
for this modification. In the MRED code at Vanderbilt \citep{Weller_bigpaper_2010_5550410},
we have been using such a technique for many years, by incorporating
conventional Geant4 hadronic processes in a cross-section biasing
wrapper. The wrapped process increases cross sections by a biasing
factor $b$, and scales the weight of tracks in which a biased process
occurs by a factor $1/b$. To first order, this is a very effective
technique, but it has limits as to how much bias can be applied. Often,
in small geometries, bias factors of a few hundred can be used, resulting
in roughly equal decreases in the computational effort required to
quantify effects involving nuclear reactions. Factors of a this order
are clearly significant, and can be critically important considering
the practical costs of computing.

There are two limiting factors which control how much one can upwardly
bias the cross section of a process in a Monte-Carlo calculation framework
such as Geant4 \citep{Agostinelli2003}. The first one, incoming beam
depletion, is addressed by this paper. If beam depletion is ignored,
there is a distortion of computed effects that is dependent upon the
amount of material an incoming particle has crossed before it interacts.
The second is innate to all types of Monte-Carlo codes; if one enhances
the probability of one class of event, it is at the cost of reducing
the statistics for some other class. Depending on the purpose of a
given calculation, one may be forced to limit the bias factor so that
events in which no reaction in the biased category occur still get
some significant sampling weight.

The problem of incoming particle depletion is, in brief, as follows.
As a particle traverses a material, it has a probability of being
deflected or destroyed by a collision with an atom in the target.
For typical nuclear cross sections, this results in an attenuation
of the incoming particle flux with a length scale of the order of
a few centimeters. This part is purely physical. However, if one enhances
the cross section for interaction by a factor $b$, the attenuation
length is unphysically reduced by the same factor. Conversely, reducing
the cross section allows a simulated particle to penetrate too deeply,
and must be compensated by reducing its statistical weight. This results
in the under-sampling or oversampling of events as one progresses
into the target. We present, in the next section, a quantitative mechanism
to account for these effects. 

The approach we present stands in contrast to methods known as particle
splitting techniques \citep{Booth_variance_reduction_1994,szieberth_non_coincidence_physor_2010}.
In those techniques, one takes each simulated particle, and when an
interaction occurs, splits the history into two branches, one that
contains the interaction products, and the other that carries away
the original particle with a modified statistical weight and allows
it to further interact. Instead of dealing with the weight adjustment
discretely, as in particle splitting, we correct for the effects of
the cross section bias in an average manner along the particle path.
This greatly simplifies the book keeping relative to particle splitting,
and allows the benefits of biasing to be incorporated into conventional
Monte-Carlo codes with almost no modification to the structure of
the code itself.

\section{Basic depletion argument}

As a particle propagates through a material, the probability $P(x)$
of it reaching a point $x$ is set by the solution of: \begin{equation}
\frac{dP(x)}{P(x)}=-\rho(x)\,\sigma(x,\, E(x))\, dx\end{equation}
where $\rho$ is the material density in scatterers/volume, $\sigma$
is the cross section for an interaction, and $x$ is the distance
travelled along the particle path, in the absence of any biasing of
the probabilities. This equation only takes account of binary interactions
of the incoming particle with the target material, and ignores spontaneous
decay of the particle. The treatment presented needs to be modified
if particle decay in flight is included; we do not treat this case
here. If we bias $\sigma$ by a factor $b$, then the probability
is modified:\begin{equation}
\frac{dP_{b}(x)}{P_{b}(x)}=-b\rho(x)\,\sigma(x,\, E(x))\, dx\end{equation}
which, noting that \begin{equation}
\frac{d(f/g)}{f/g}=\frac{df}{f}-\frac{dg}{g}\end{equation}
 results in an extra depletion of the probability of it penetrating
to $x$ of\begin{equation}
\frac{d\left(P_{b}(x)/P(x)\right)}{P_{b}(x)/P(x)}=-(b-1)\rho(x)\,\sigma(x,\, E(x))\, dx\label{eq:first-bias-correction}\end{equation}

If one therefore gradually adjusts the statistical weight $W$ of
a particle according to:\begin{equation}
\frac{dW(x)}{W(x)}=(b-1)\rho(x)\,\sigma(x,\, E(x))\, dx\end{equation}
then, the weighted probability $P_{b}(x)W(x)$ will be equal to the
original probability \textbf{$P(x)$}. This effectively conserves
particle flux, albeit by carrying a smaller number of more heavily
weighted particles (in the case of $b>1)$. 

In typical simulations, one takes discrete steps over which the parameters
are held constant. At the end of such a simulation step, where $\rho$
and $\sigma$ are taken to be constant,

\begin{multline}
\frac{\Delta W(x)}{W(x)}=(b-1)\rho(x)\,\sigma(x,\, E(x))\,\Delta x\\
\equiv(b-1)\times\Delta\lambda\end{multline}
where $\Delta\lambda$ is the number of physical interaction lengths
traversed in this step. The solution to this is:

\begin{equation}
W(x)=W(x-\Delta x)\times\exp\left((b-1)\times\Delta\lambda\right)\label{eq:full_correction}\end{equation}

The effect of applying this correction is that the weight of particles
is adjusted in an average sense. Instead of tracking each individual
step of each particle, and adjusting the weight and splitting particle
histories so that probability is exactly conserved depending upon
the events that befall a particular particle, this approach corrects
the weight so that the average effect of this correction compensates
the average error due to the adjustment of the strength of an interaction.
Since, from equation \ref{eq:first-bias-correction}, we know the
exact value of the correction, this approach is statistically exact.
That is all that is needed for a calculation to produce an unbiased
result using a statistical method such as Monte Carlo integration. 

As will be seen graphically in the examples below, the selection of
the bias parameter for a given problem is not too critical. If the
bias parameter is far too small (in the case of thin target problems,
where it is greater than unity), many of the particles penetrate the
target without interacting, resulting in lower computational efficiency.
If the bias parameter is far too large (only a small fraction of the
particles not interacting), the statistical quality of points deep
in the target will be degraded due to beam depletion, although the
computed mean value will still be unchanged. The converse is true
for thick-target situations, where one is reducing the cross section
to permit penetration, but the outcome is the same. As a result, it
is generally very efficient to choose a bias such that roughly half
of the particles are absorbed; this achieves a balance between the
number of particles that interact and the statistical significance
for particles that have penetrated far into the target.

\section{Procedure for applying the bias correction}

In a typical Monte-Carlo particle transport code, particles interact
with their environment through two classes of processes: 
\begin{enumerate}
\item Discrete processes, which represent point-like interactions of the
particle with another particle in its proximity, and,
\item Continuous processes, which represent smooth drag-like terms which
occur along the entire length of each step in a particle's trajectory.
\end{enumerate}
Also, it is assumed a particle carries along with it a statistical
weight $w$ that can be modified as it propagates, and that can be
passed along as a starting weight to any daughter particles created.

The method we describe is specifically applicable to the discrete
processes. The procedure for transporting a particle along one step
of a trajectory is roughly as follows:
\begin{enumerate}
\item Start with a particle at position $\vec{x_{0}}$ with energy $E_{0}$
\item Compute a transport length $L_{\mathrm{trans}}$ for the next step
the particle will take, based on the summed strength of the interaction
processes computed at position $\vec{x}_{0}$ and energy $E_{0}$. 
\item \label{enu:end-of-step-process}Randomly select which of multiple
discrete interaction processes will be invoked at the end of the step,
based on the individual strengths of the applicable processes.
\item Move particle forward a distance $L_{\mathrm{trans}}$ along its current
momentum direction.
\item Compute the effects which all applicable continuous interactions would
have had on the particle, and update its parameters based on those
effects.
\item Multiply the particle's statistical weight by \begin{equation}
\prod_{i=1}^{n}\exp\left(\left(b_{i}-1\right)\times L_{\mathrm{trans}}\times N\times\sigma_{i}\right)\label{eq:along_step_accum_weight_adjust}\end{equation}
 where $i$ indexes over the active discrete processes, $b_{i}$ and
$\sigma_{i}$ are the bias in effect and the cross section for the
$i^{th}$ process at this step, and $N$ is the material particle
density along the step such that $L_{\mathrm{trans}}N\sigma=\Delta\lambda$
as defined in equation \ref{eq:full_correction}.
\item Multiply the particle's statistical weight by $1/b_{j}$ where the
index $j$ is that associated with the process chosen in step \ref{enu:end-of-step-process}.
\item Invoke the discrete interaction selected in step \ref{enu:end-of-step-process},
and update the particle's parameters. Create daughter particles with
the current statistical weight of the parent.
\end{enumerate}

\section{Management of statistical weights}

At the end of tracking of all the particles that are created during
a simulation event, one typically accumulates the information about
how energy was deposited by those particles. We define the following
quantities, assuming the particle indexed by $k$ in the event starts
out with weight $w_{k,0}$.

The statistical weight of the $k^{th}$ particle at the end of the
$n^{th}$ step along its trajectory is:\begin{equation}
w_{k,n}=w_{k,0}\prod_{l=1}^{N{}_{\mathrm{b}\, k,n}}\frac{1}{B_{k,l}}\prod_{m=1}^{n}\prod_{i=1}^{N{}_{\mathrm{p}\, k}}\exp\left(\left(b_{k,m,i}-1\right)\times\Delta\lambda_{k,m,i}\right)\label{eq:weight_product}\end{equation}
where the terms are as defined in eq. \ref{eq:full_correction} except
that $b_{k,m,i}$ and $\lambda_{k,m,i}$ refer specifically to the
bias and scaled interaction length that were applicable to the $i^{th}$
process along the $m^{th}$ step, $N{}_{\mathrm{p}\, k}$ is the number
of processes for the $k^{th}$ particle, $B_{k,l}$ is the $l^{th}$
discrete bias applied to the $k^{th}$ particle, and $N{}_{\mathrm{b}\, k,n}$
is the number of discrete interactions with a bias factor different
than $1$ that have been applied to the $k^{th}$ particle prior to
and including the $n^{th}$ step. This is just the accumulated result
of eq. \ref{eq:along_step_accum_weight_adjust} along all the steps,
including multiple processes creating biases. If a particle interacts
at step $n$ and produces daughters, the daughter tracks are created
with a weight $w_{k,n}$ of the parent particle $k$. 

Also, define a weight for an entire event,

\begin{equation}
W_{\mathrm{e}}=w_{0}\,\prod_{l=1}^{N_{\mathrm{int}}}\frac{1}{B_{l}}\prod_{k=1}^{N_{\mathrm{part}}}\prod_{m=1}^{N_{\mathrm{steps}\, k}}\prod_{i=1}^{N_{\mathrm{p}\, m}}\exp\left(\left(b_{k,m,i}-1\right)\times\Delta\lambda_{k,m,i}\right)\label{eq:event-weight_product}\end{equation}
where, in this case, it is assumed the event started with a statistical
weight $w_{0}$, $N_{\mathrm{int}}$ discrete interactions occurred
in the course of the event, each biased by a factor $B_{l}$, $N_{\mathrm{part}}$
is the number of particles in the entire event, $N_{\mathrm{steps}\, k}$
is the number of steps for the $k^{th}$ particle, and the inner product
runs over all segments of all tracks in the event, with \textbf{$b_{k,m,i}$}
and $\lambda_{k,m,i}$ as in eq. \ref{eq:weight_product}. This accumulates
all of the weight modifications from the event. Note that the book-keeping
here creates a result that is not simply the products of the weights
of all the particles in the event at the end of their trajectory.
Since an interaction may produce multiple daughters, the product of
all the final weights contains replicates of the bias that is introduced
by an interaction. Eq. \ref{eq:event-weight_product} counts each
bias exactly once.

It is necessary to consider the meaning of the statistical weights
that can be computed by this process. This is a somewhat complex issue,
because, in different use cases, it may be necessary to treat the
tracks in the event as individual entities, each with a distinct statistical
weight, or to treat the entirety of the event as a single ensemble
with one global weight. The difference between these two approaches
depends on assumptions of correlated or uncorrelated statistical probabilities.
The two cases just described are the limiting cases, and result in
easy-to-interpret statistics. More complex cases could be considered,
in which one wishes to calculate the joint probability of two arbitrary
tracks in an event. In this case, it is necessary to look at their
parent lineage, and compute the statistics with the knowledge that
when they share a vertex, the statistical weight change resulting
from that shared vertex is fully correlated, while changes which occur
after they no longer share a vertex are uncorrelated. We do not consider
this in the discussion below, because it is not relevant to the most
common use cases.

The weight obtained from eq. \ref{eq:event-weight_product} is the
effective statistical weight for all the correlated components which
have happened in an event. If the event is to be treated as a whole,
single object, as in non-Boltzmann tallies such as energy depositions
in sensitive detectors being entered into a histogram as pulse heights,
or as contributions to a coincidence rate, all tracks of the event
should be treated as having this single weight, and the individual
weights on the tracks should be ignored. In the typical format of
a call to a histogramming tool such as \noun{AIDA} \citep{AIDA_main,OpenScientist},
one would execute a call such as \noun{fill()} to add an event of
amplitude $E_{\mathrm{dep}}$ with weight \noun{$W_{\mathrm{e}}$}
to accumulate the result from a sensitive detector at the end of an
event. 

On the other hand, if one is computing total dose (for example), where
each track and step in the event is treated independently and one
is computing an integral across many events, the weights which are
left attached to the individual tracks are the appropriate weights
to use. Each particle in the event has a specific probability of being
created by its parent, and this is properly accounted for by eq. \ref{eq:weight_product}.
The appropriate mechanism for adding energy into a calorimeter is
to add \noun{$E_{\mathrm{dep}\, k,n}\times w_{k,n}$}, where $w_{k,n}$
is as defined in eq. \ref{eq:weight_product} and $E_{\mathrm{dep}\, k,n}$
is the energy deposited by the $k^{th}$ particle during the $n^{th}$
step.

\section{Specific Implementation in Geant4}

In this section, we describe the details of the implementation of
this process in Geant4. The purpose is both for the benefit of Geant4
users, and also to present code which is an actual implementation
of eqs. \ref{eq:along_step_accum_weight_adjust} and \ref{eq:event-weight_product},
since the code may be easier to interpret or reuse than the formulas
themselves.

This code is written as a wrapper for \noun{G4VDiscrete\-Process}
in Geant4 (for information on all Geant4 specific information, see
\citep{Geant4Developers}), and exposes itself as a \noun{G4V\-Continuous\-Discrete\-Process},
so that it can provide an \noun{AlongStepDoit()} method to carry out
the bias adjustment described in equation \ref{eq:full_correction}.
It shares the bias factor among all instances of the wrapper, so that
one can adjust all particles with a single call, without having to
keep a registry of copies of this process, and adjust every one when
the bias is changed. As discussed in the shielding example in section
\ref{sec:Samples}, there might be use cases for which there would
be benefit in biasing different processes by different amounts. This
should be easily adapted to that case. 

The mechanism for using this process is as follows. For each \noun{G4V\-Discrete\-Process}
to be biased, one creates an instance of the wrapper with the already-created
process to be wrapped as the argument to the constructor. This can
either be done when the wrapped processes are originally created (in
the physics list construction), or afterwards. To do it afterwards
requires scanning the process tables for all particles for candidates
to be wrapped, and replacing the process table entries with newly
instantiated copies of the wrapper. This allows one to use the built-in
Geant4 standard physics lists, and then apply the wrappers as an independent
step. The results shown in section \ref{sec:Samples} were created
using biased processes in \noun{Hadron\-Physics\-QGSP\_BIC\_HP}
for the inelastic physics, and unbiased \noun{G4UHadron\-Elastic\-Process}
attached to the \noun{G4Neutron\-HPorL\-Elastic} model for the elastic
scattering. 

To interface with the event-based weighting created by this process
requires code in the \noun{G4User\-Event\-Action} class, with a
call to \noun{Reset\-Accumulated\-Weight()} in the \noun{Begin\-Of\-Event\-Action}().
This resets the accumulated event weight to unity. Then, in the \noun{End\-Of\-Event\-Action()}
or in \noun{G4V\-Sensitive\-Detector} \noun{End\-Of\-Event}()
processing, one calls \noun{Get\-Accumulated\-Weight()} to retrieve
the appropriate statistical weight for the event, as defined by eq.
\ref{eq:event-weight_product}. 

On the other hand, if one is computing total dose (for example), the
appropriate mechanism for adding energy into a calorimeter is to add
\noun{deposited\_energy$\times$track}.\noun{Get\-Track\-Weight}()
(where the weight at the $n^{th}$ step is as defined in eq. \ref{eq:weight_product})
for each hit on a sensitive detector during the ongoing processing
of an event. If this approach is being used, there is no need (although
no harm) to call \noun{Reset\-Accumulated\-Weight()}.

\section{Sample runs showing the effects of biasing\label{sec:Samples}}

In this section, we present examples, based on a few archetypal models.
In the first set of examples, the system we use as our exemplar of
increased bias is a $\unit[1]{mm}$ film of lightly borated $\left(\mathrm{CH}_{2}\right)_{n}$
(essentially polyethylene or polypropylene). It is irradiated with
a monoenergetic neutron beam, with an energy of $\unit[0.025]{eV}$.
The boron concentration is $\unit[0.1]{wt.}\%$ in some cases, and
$\unit[0.2]{wt.}\%$ in others (indicated in each example). This results
in weak depletion of the incoming beam in reality. The alpha particle
production yield is calculated in 10 bins throughout the thickness
of the film. The second case, of strong actual attenuation in a shielding
computation, is best carried out with a bias of less than unity. In
the final example, we present a coincidence rate calculation in a
small target typical of semiconductor problems, demonstrating the
ability of this technique in non-Boltzmann problems. 

\begin{figure}
\noindent \centering{}\includegraphics[width=1\columnwidth]{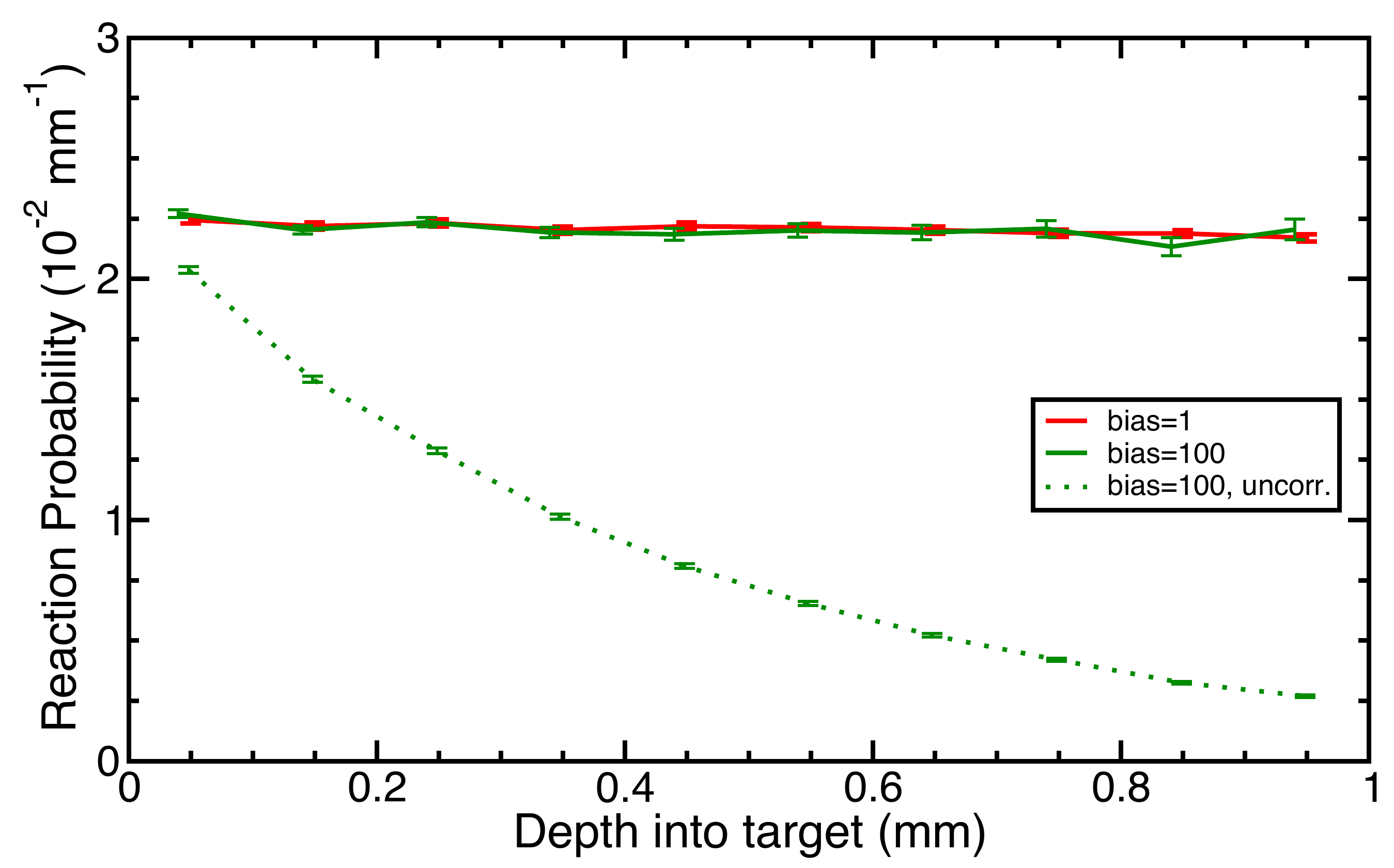}\caption{\label{gold_1e-3}Comparison of $\alpha$ particle yield vs. depth
in a $\unit[0.1]{wt.}\%$ boron loaded $\left(\mathrm{CH}_{2}\right)_{n}$
film with $\unit[0.025]{eV}$ incident neutrons. The bias=1 curve
required 100 times as many events ($10^{7}$), and 25 times the computer
time, as the bias=100 curve, for approximately the same statistical
quality. Points have in the upper group of curves have been slightly
shifted horizontally to avoid overlaps. Curves in this figure and
figures below labeled 'uncorr.' are for runs with biased cross sections,
but do not include the weight correction of eq. \ref{eq:full_correction};
they are included since they show the level of primary beam depletion,
which is an important measure of the strength of the biasing.}

\end{figure}

\subsection*{High-statistics accuracy demonstration}

Figure \ref{gold_1e-3} shows a yield comparison between a run with
$10^{7}$ particles with a bias of 1, and $10^{5}$ particles with
a bias of 100. Since the absorption is weak, most of the neutrons
penetrate the target without interacting in the unbiased case. The
computational time for the unbiased case is 25 times greater than
that for the biased case, even though the number of particles run
is 100 times greater. This discrepancy occurs because events that
include an interaction take more time than those in which there is
no interaction. Fig. \ref{gold_1e-3} demonstrates three important
points. First, the biased computation returns the same result as the
unbiased one throughout the target thickness, even though there is
significant depletion of the primary particles in the biased case.
Second, because of increased computation associated with the larger
number of reactions in the biased case, the efficiency of the computation
increases sub-linearly with the bias, and would saturate when essentially
all of the particles are interacting. Third, close examination of
the plot shows that the uncertainty in the yield for the biased case
increases with depth. This is because the increased depletion results
in poorer statistics for particles that have traversed a larger distance.
This effect also sets an upper limit on the bias that can be applied.

\begin{figure}
\noindent \centering{}\includegraphics[width=1\columnwidth]{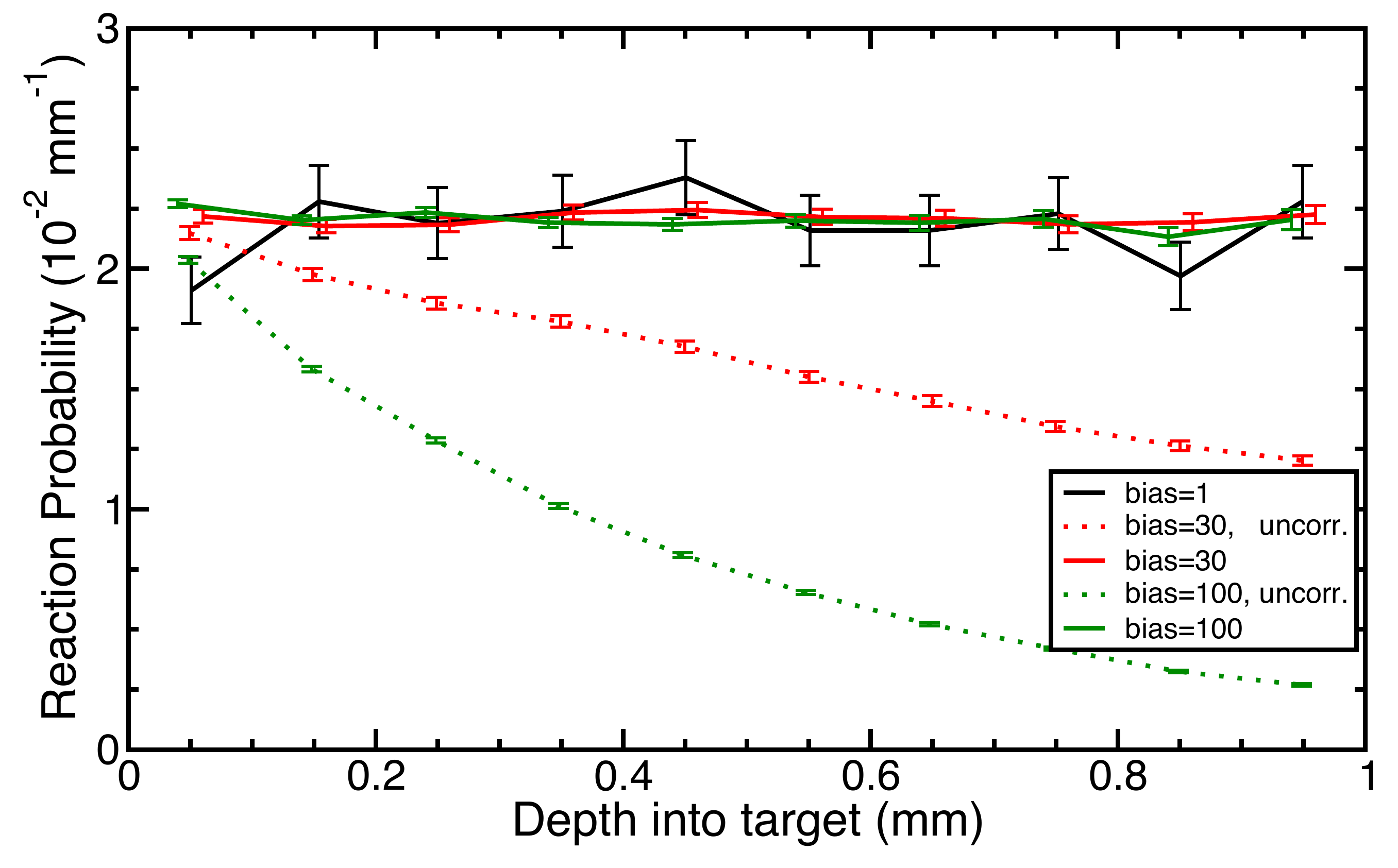}\caption{\label{bias_1_30_100}Comparison of $\alpha$ particle yield vs. depth
in a $\unit[0.1]{wt.}\%$ boron loaded $\left(\mathrm{CH}_{2}\right)_{n}$
film with $\unit[0.025]{eV}$ incident neutrons. All of these data
sets were run with the same number of events ($10^{5}$). Note that,
in this example, the statistical uncertainties are fairly uniformly
distributed for biases up to about 30. Points in the upper group of
curves have been slightly shifted horizontally to avoid overlaps.
See the fig. \ref{gold_1e-3} caption for discussion of 'uncorr.'
curves.}

\end{figure}

\subsection*{Moderately attenuated transport (bias well chosen)}

Figure \ref{bias_1_30_100} shows a yield comparison made with $10^{5}$
particles in each run, using different bias factors, again on the
$\unit[0.1]{wt.}\%$ target. The feature to observe in this result
is that one can tune the distribution of errors by adjusting the bias,
and that by turning the bias too high, the variance is increased for
particles that are arriving in a region where the beam has been heavily
depleted. In this case, although the result for a bias of 100 has
smaller statistical errors at the entrance to the target than those
for the bias of 30, it has larger errors at the exit. This behavior
gives guidance about reasonable bias factors to use. In general, there
is little to gain in a calculation if the bias is set so large that
more than roughly half of the beam is depleted. 

\begin{figure}
\noindent \centering{}\includegraphics[width=1\columnwidth]{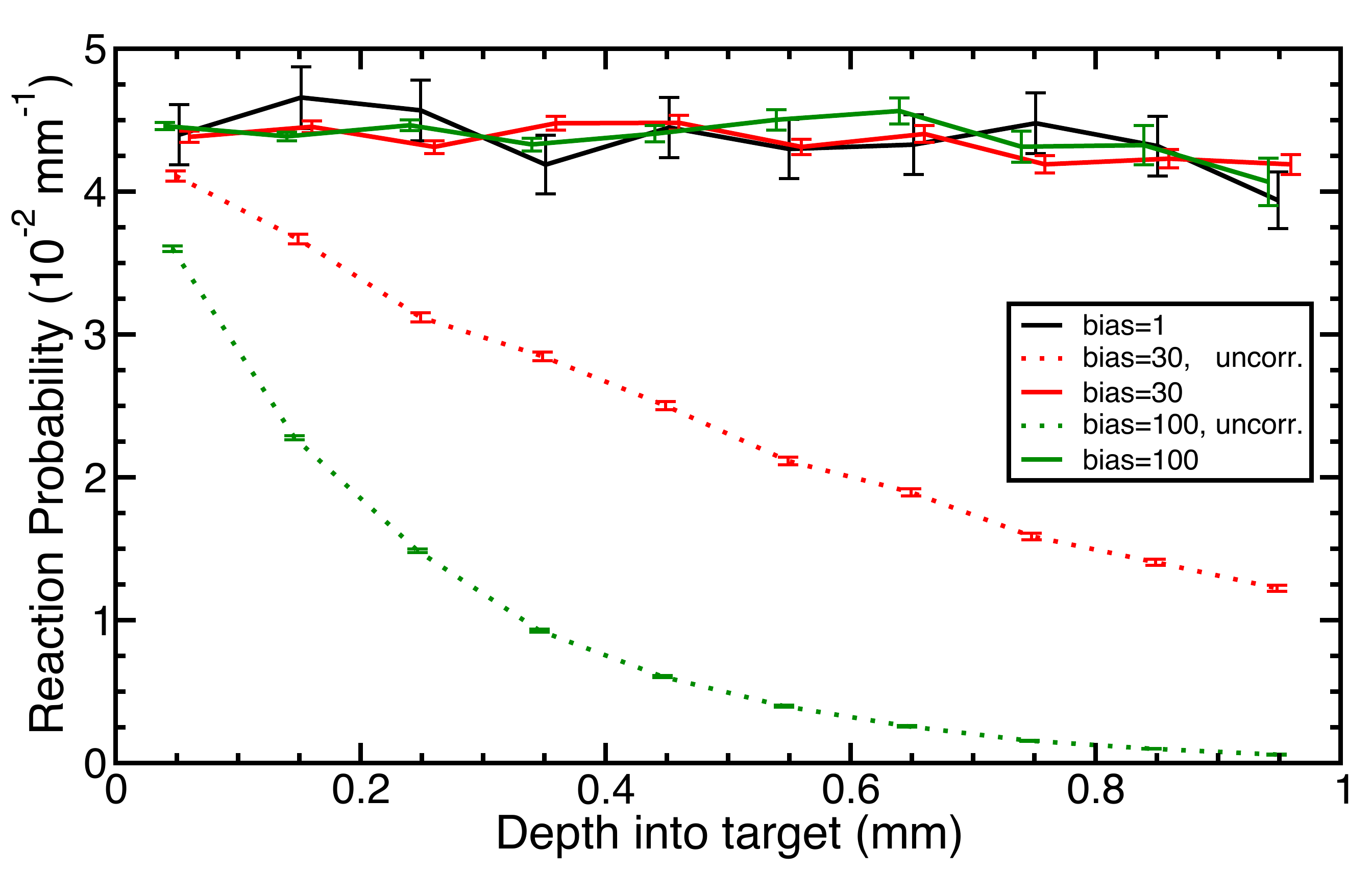}\caption{\label{max_depletion}Comparison of $\alpha$ particle yield vs. depth
in a $\unit[0.2]{wt.}\%$ boron loaded $\left(\mathrm{CH}_{2}\right)_{n}$
film with $\unit[0.025]{eV}$ incident neutrons. In this case, the
higher boron concentration results in extreme beam depletion for the
bias 100 run, and greatly increased statistical uncertainties of the
yield for the points deepest into the target. Points in the upper
group of curves have been slightly shifted horizontally to avoid overlaps.
See the fig. \ref{gold_1e-3} caption for discussion of 'uncorr.'
curves.}

\end{figure}

\subsection*{Strongly attenuated transport (bias too large)}

Figure \ref{max_depletion} shows the result of extreme depletion
in the incoming beam, due to both strong absorption and high bias.
In this case, the boron concentration in the target was raised to
$\unit[0.2]{wt.}\%$. Since the transmission is exponential in the
concentration, very little of the incoming beam is transmitted, and
the statistical uncertainties for the highest bias are very large
at the exit of the target. Although the result of the calculation
still has the correct mean, its variance is very large.

\begin{figure}
\noindent \centering{}\includegraphics[width=1\columnwidth]{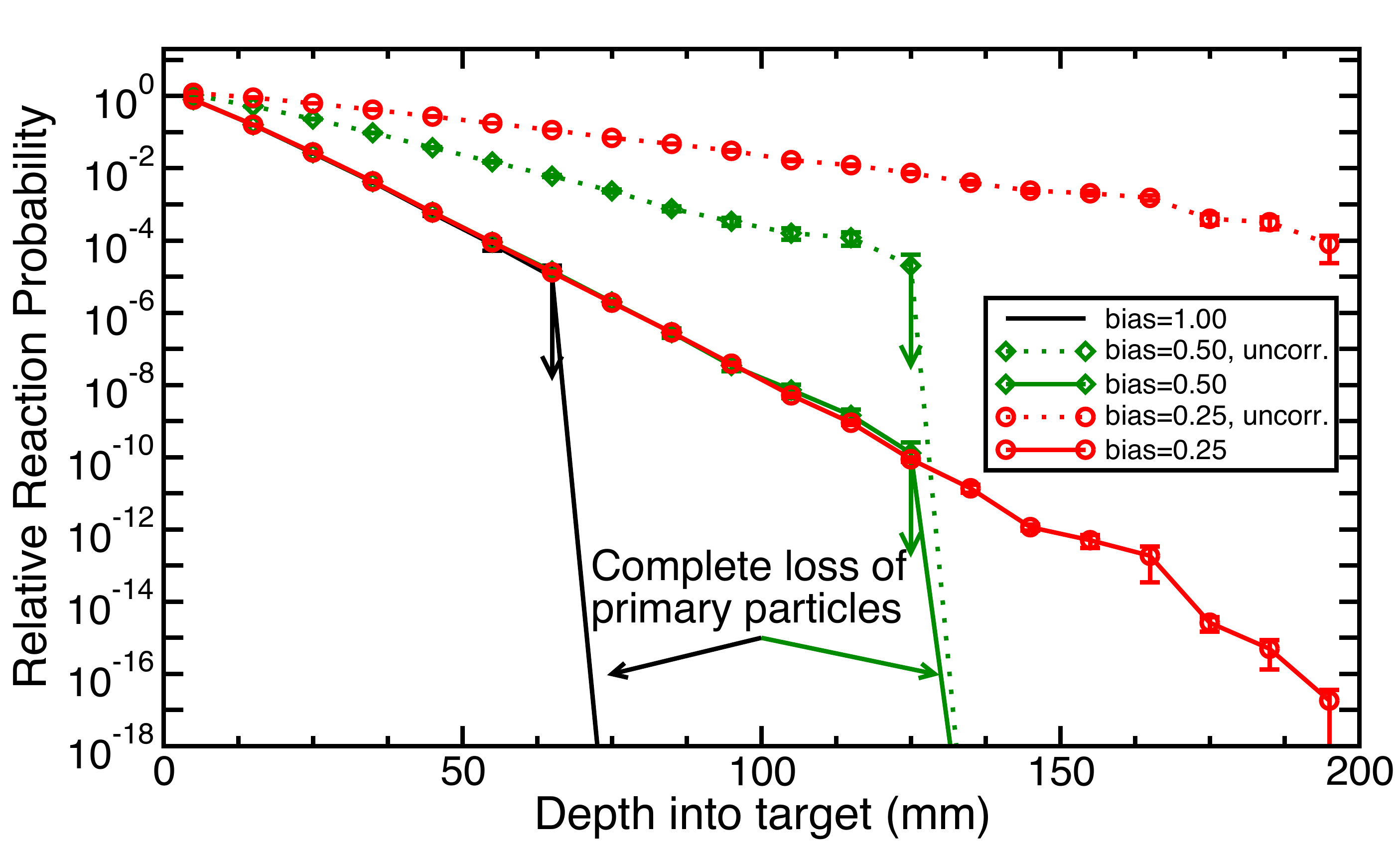}\caption{\label{shielding}Comparison of $\alpha$ particle yield vs. depth
in a $\unit[5]{wt.}\%$ boron loaded $\left(\mathrm{CH}_{2}\right)_{n}$
film with a beam of $\unit[2]{eV}$ incident neutrons. Note that with
bias less than unity, a shielding calculation can be carried out involving
$10^{15}$ attenuation, using $10^{5}$ incident particles (red curve).
See the fig. \ref{gold_1e-3} caption for discussion of 'uncorr.'
curves.}

\end{figure}

\subsection*{Shielding calculation (bias less than unity)}

Figure \ref{shielding} demonstrates the results of using the same
biasing scheme, but with a $b$ value less than unity for both the
inelastic and elastic processes. This demonstrates the important ability
to use this biasing mechanism in cases in which the true primary beam
depletion is so large that physics deep inside a target is poorly
sampled. This case arises frequently in the computation of leakage
of shielding systems, which can become very computationally expensive
to carry out without variance reduction. This is equivalent to the
{}``forced-flight'' technique, such as is provided in the MCBEND
code \citep{shuttleworth_mcbend_1999,Wright2004162}, with the added
flexibility that one can both calculate the attenuation of the primary
particle flux, and the production of correctly-weighted secondaries.

\begin{figure}
\noindent \begin{centering}
\includegraphics[width=0.9\columnwidth]{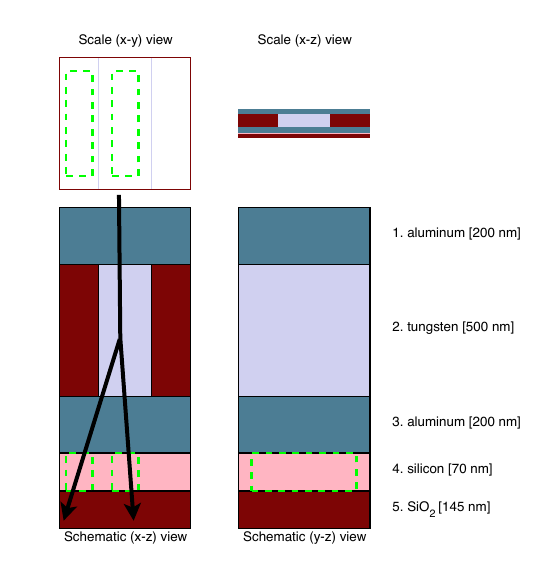}\caption{\label{fig:Multilayer-stack}Multilayer stack of materials, with two
sensitive volumes (marked in dashed green lines). Coincidences were
computed in this structure when it was irradiated from above with
6.4 GeV alpha particles. The heavy black arrows show a typical nuclear
reaction event that could create a coincidence.}

\par\end{centering}

\end{figure}

\subsection*{Coincidences (non-Boltzmann problem)}

A final important case is a test of this weighting system in an explicitly
non-Boltzmann transport problem. The archetype we use for this is
a computation of a coincidence cross section for energy deposition
in two volumes separated in space. In our test geometry, energy can
be deposited in coincidence by nuclear reactions that include multiple
particles in the outgoing channel. In figure \ref{fig:Multilayer-stack}
we show the target that was used to test the rate of coincidence generation,
using the weighting in eq. \ref{eq:event-weight_product}. This target
was exposed to $10^{5}$ $\unit[6.4]{GeV}$ alpha particles at normal
incidence to the top, with a bias factor $b=10^{4}$, using the CREME
Monte-Carlo tool \citep{creme_site}. A coincidence is determined
to be an energy deposition of more than $\unit[10]{keV}$ in each
of the sensitive volumes (shown as dotted green boxes in layer 4 of
figure \ref{fig:Multilayer-stack}). For comparison, we also computed
the coincidence cross section with unbiased statistics, running $10^{9}$
particles. The result was $\unit[(2.0\pm0.2)\times10^{-14}]{cm^{2}}$
for the biased case and $\unit[(2.2\pm0.2)\times10^{-14}]{cm^{2}}$
for the unbiased case. The computational time required for the biased
case was 2 minutes; for the unbiased case, it was 27 hours.

\section{Conclusion}

This work presents a very simple mechanism for the adjustment of the
effective strength of an interaction that can be described by a cross
section, in a way that precisely preserves the mean transport characteristics,
while allowing strong variance reduction. It allows one to efficiently
calculate particle transport and scattering through solids in the
extreme cases of either very weak or very strong scattering by the
material, either of which can result in very large computational demands
to achieve a low-variance result throughout the target material. 

For targets that produce very little scattering of the incoming particles,
the effective strength of the interaction can be increased, resulting
in fewer particles that pass through the target un-scattered, and
therefore less lost computation effort on particles that do not contribute
to the final result. Conversely, for targets in which, in reality,
most of the incoming particles would be absorbed before they penetrate
to the depth of interest in the target, the computed interaction can
be weakened to transport more particles to a greater depth into the
target, and reduce the variance of results computed there. 

In either case, the continuous adjustment of the weight of un-scattered
particles combined with the specified computation of the total event
weight assures that the computed transport properties remain unaffected
by the introduction of the cross-section biasing factor. This holds
true for both Boltzmann and non-Boltzmann type problems.

\section*{Acknowledgements}

Work supported by the Defense Threat Reduction Agency (DTRA) Basic
Research Program. Additional support was provided by the NASA Electronic
Parts and Packaging program (NEPP).\bibliographystyle{unsrtnat}
\bibliography{Mendenhall,RadiationEffects}

\onecolumn

\section*{Appendix: Sample Geant4 implementation}

The purpose for including an appendix in this paper with the source
code for a complete Geant4 wrapper process that implements the above
algorithm is twofold. First, the code is an alternative, iterative
description of the deeply nested products in equations \ref{eq:weight_product}
and \ref{eq:event-weight_product}, and as such may prove more readable
than the equations themselves, especially as they may be implemented
in other codes. As documentation of the algorithm, it should not be
too difficult to read around the Geant4-specific naming and structure
to get to the functionality, which is almost entirely in the main
code body. Second, the code should be able to be compiled and directly
used in Geant4 simulations.

\section*{Header file: GroupedSigmaBooster.h}

\inputencoding{latin9}
\begin{lstlisting}[language={C++},showstringspaces=false,tabsize=2]
#ifndef GroupedSigmaBooster_h
#define GroupedSigmaBooster_h 1

#include "globals.hh"
#include "G4VProcess.hh"
#include "G4VContinuousDiscreteProcess.hh"
#include "G4VDiscreteProcess.hh"
#include "G4Track.hh"

class GroupedSigmaBooster : public G4VContinuousDiscreteProcess
{
public:
	// construct the booster as a wrapper of the provided process.
	// this will fail if the process is not a G4VDiscreteProcess.
	GroupedSigmaBooster(
		const G4String& processName = "SigmaBooster",
		G4VProcess *wrapProc=0):
			G4VContinuousDiscreteProcess(processName+":"+wrapProc->GetProcessName()),
			lastStepBoost(1.0) 
	{
		proc=dynamic_cast<G4VDiscreteProcess *>(wrapProc);
		if (!proc) 
			G4Exception("Cannot wrap a non-G4VDiscreteProcess" 
				"with sigma booster: "+wrapProc->GetProcessName());
		pParticleChange=new G4ParticleChange;
	}

	~GroupedSigmaBooster() { 
		if(pParticleChange) delete pParticleChange;  
		pParticleChange=0;
	}

	virtual G4double PostStepGetPhysicalInteractionLength(
		  const G4Track& aTrack, G4double   prev, G4ForceCondition* cond);
	
	virtual G4double AlongStepGetPhysicalInteractionLength(
			const G4Track&, G4double  , 
			G4double  , G4double& , G4GPILSelection* ) 
		{ return DBL_MAX; }

    virtual G4double GetContinuousStepLimit(const G4Track& ,
			G4double  , G4double  , G4double&  )
		{ return DBL_MAX; }

	virtual G4VParticleChange* PostStepDoIt(
		const G4Track& aTrack, const G4Step& aStep);
	virtual G4VParticleChange* AlongStepDoIt(
		const G4Track& aTrack, const G4Step& aStep);
	
	// pass through most of the virtual calls to the underlying process
	virtual void  ResetNumberOfInteractionLengthLeft() 
	{
		lastStepBoost=1.0;
		G4VContinuousDiscreteProcess::
			ResetNumberOfInteractionLengthLeft();
		proc->ResetNumberOfInteractionLengthLeft(); 
	}

	virtual void StartTracking(G4Track*aTrack) 
	{
		G4VContinuousDiscreteProcess::StartTracking(aTrack);
		proc->StartTracking(aTrack);
	}

	virtual void EndTracking() {
		G4VContinuousDiscreteProcess::EndTracking();
		proc->EndTracking();
	}

	virtual void SetProcessManager(const G4ProcessManager* procMan) 
		{  proc->SetProcessManager(procMan); } 
	virtual  const G4ProcessManager* GetProcessManager()
		{ return proc->GetProcessManager(); }
	virtual G4bool IsApplicable(const G4ParticleDefinition& aParticleType) 
		{ return proc->IsApplicable(aParticleType); }
	virtual void BuildPhysicsTable(const G4ParticleDefinition& aParticleType) 
		{ proc->BuildPhysicsTable(aParticleType); }
	virtual void PreparePhysicsTable(const G4ParticleDefinition& aParticleType) 
		{ proc->PreparePhysicsTable(aParticleType); }
	virtual G4bool StorePhysicsTable(const G4ParticleDefinition* ptcl,
		const G4String& fname, G4bool ascii = false) 
		{ return proc->StorePhysicsTable(ptcl, fname, ascii); }
	virtual G4bool RetrievePhysicsTable( const G4ParticleDefinition* ptcl,
		const G4String& fname, G4bool ascii = false) 
		{ return proc->RetrievePhysicsTable(ptcl, fname, ascii); }

	// set the bias factor for all the processes wrapped by any instance of this class.
	static void SetCrossSectionBias(G4double BiasFactor=1) 
		{ biasFactor=BiasFactor; }
	
	// if PrimaryOnly is set, only the primary track gets cross sections boosted
	// this is the default mode
	static void SetPrimaryOnly(G4bool flag) 
		{ primary_only=flag;  } 

	// control whether weights are used.  
	// This is almost never changed from its default value of 'true'
	static void SetUseWeighting(G4bool flag) 
		{ useWeighting=flag; } 
	
	// enable use of continuous correction for biasing.
	// this is almost never changed form its default of 'true'
	static void SetEnableProbabilityConservation(bool flag)
		{ conserve_probability=flag; }
	
	// get our wrapped process
	G4VDiscreteProcess *GetProcess() {return proc;}

	// get the product of all the boosts applied since weight was reset
	static G4double GetAccumulatedWeight() 
		{ return accumulatedWeight; }

	// reset accumulated weight to 1, and enable accumulation
	static void ResetAccumulatedWeight() 
		{ accumulatedWeight=1; weightInitialized=true; }
	
protected:	
		virtual G4double GetMeanFreePath(
			 const G4Track& aTrack, G4double   prev, G4ForceCondition* cond) 
		{
			// this should never get called!  It is handled through GetPhysical...
			G4Exception("Booster GetMeanFreepath called, should not happen!\n");
		}
	
	G4VDiscreteProcess *proc;
	G4double lastStepBoost, lastInteractionLengths;
	
	static G4double biasFactor;
	static G4bool primary_only, useWeighting, conserve_probability;

	static G4double accumulatedWeight;
	static G4bool weightInitialized;
	
};
#endif
\end{lstlisting}
\inputencoding{utf8}

\section*{Main code body: GroupedSigmaBooster.cc}

\inputencoding{latin9}
\begin{lstlisting}[language={C++},showstringspaces=false,tabsize=2]
#include "GroupedSigmaBooster.h"
G4double GroupedSigmaBooster::biasFactor=1.0;
G4bool GroupedSigmaBooster::primary_only=1;
G4bool GroupedSigmaBooster::useWeighting=0;
G4bool GroupedSigmaBooster::conserve_probability=true; 
G4double GroupedSigmaBooster::accumulatedWeight=0;
G4bool GroupedSigmaBooster::weightInitialized=false;

G4double GroupedSigmaBooster::PostStepGetPhysicalInteractionLength(
	const G4Track& aTrack, G4double   prev, G4ForceCondition* cond) 
{
	if(biasFactor == 0.0) { // special case, turn process off completely 
		lastStepBoost=0.0;
		return DBL_MAX; // and do it via the PIL, too.
	}
	G4double pil=proc->PostStepGetPhysicalInteractionLength(
		aTrack, prev*lastStepBoost, cond);
	
	if( !primary_only || aTrack.GetTrackID()==1) {
		pil /= biasFactor;
		lastStepBoost=biasFactor;
	} else lastStepBoost=1.0;
		
	return pil;
}

G4VParticleChange* GroupedSigmaBooster::AlongStepDoIt(
	const G4Track& aTrack, const G4Step& aStep) 
{
	pParticleChange->Initialize(aTrack);

	if(conserve_probability && lastStepBoost != 1.0) {
		// this step didn't interact, so we boost weight appropriately
		G4ParticleChange* aParticleChange=
			static_cast<G4ParticleChange *>(pParticleChange);
		aParticleChange->SetParentWeightByProcess(false); 
		// SetParentWeightByProcess allows us to change the weight
		G4double adjust=
			std::exp((lastStepBoost-1)*
				(aStep.GetStepLength()/proc->GetCurrentInteractionLength() )
			);
		G4double newWeight=aTrack.GetWeight()*adjust;
		aParticleChange->ProposeWeight(newWeight); // up-weight primary
		if(weightInitialized) accumulatedWeight*=adjust;
	}	
	return pParticleChange;
}

G4VParticleChange* GroupedSigmaBooster::PostStepDoIt(
	const G4Track& aTrack, const G4Step& aStep) 
{
	G4VParticleChange* vParticleChange=proc->PostStepDoIt(aTrack, aStep);
	G4ParticleChange* aParticleChange;
	if(useWeighting && lastStepBoost != 1.0 && 
	   (aParticleChange=dynamic_cast<G4ParticleChange *>(vParticleChange)) != 0 ) 
	{
		// only do dynamic cast if all other tests pass, 
		// since it is slightly expensive
		int nSec=aParticleChange->GetNumberOfSecondaries();
		G4double newWeight=aTrack.GetWeight()/lastStepBoost;		
		// set weights of secondaries down 
		for(int i=0; i<nSec; i++) 
			aParticleChange->GetSecondary(i)->SetWeight(newWeight);
		aParticleChange->SetParentWeightByProcess(false); 
		aParticleChange->ProposeWeight(newWeight); // down-weight primary
	
		if(weightInitialized) accumulatedWeight/=lastStepBoost;

	}
	return vParticleChange;
}
\end{lstlisting}
\inputencoding{utf8}
\end{document}